\documentclass[conference,10pt]{IEEEtran}
\usepackage[utf8]{inputenc}
\def\BibTeX{{\rm B\kern-.05em{\sc i\kern-.025em b}\kern-.08em
    T\kern-.1667em\lower.7ex\hbox{E}\kern-.125emX}}

\usepackage{graphicx}
\usepackage{color}
\usepackage{xcolor}
\usepackage{cite}
\usepackage{amsmath}
\usepackage[caption=false]{subfig}
\usepackage{amssymb}
\usepackage{comment}
\usepackage{mathtools}
\usepackage{tabulary}
\usepackage{acronym}

\newacro{ANN}[ANN]{Artificial Neural Network}
\acrodefplural{ANN}[ANNs]{Artificial Neural Networks}

\newacro{RIS}[RIS]{Reconfigurable Intelligent Surface}
\acrodefplural{RIS}[RISs]{Reconfigurable Intelligent Surfaces}

\newacro{6G}[6G]{Sixth Generation}
\newacro{SNR}[SNR]{Signal-to-Noise Ratio}
\newacro{SINR}[SINR]{Signal-to-Interference-plus-Noise Ratio}
\newacro{TX}[TX]{Transmitter}

\newacro{RX}[RX]{Receiver}
\acrodefplural{RX}[RXs]{Receivers} 

\newacro{TVC}[TVC]{Time Varying Channel}
\acrodefplural{TVC}[TVCs]{Time Varying Channels}

\newacro{BS}[BS]{Base Station}
\newacro{UE}[UE]{User Equipment}
\acrodefplural{UE}[UEs]{Users Equipments}

\newacro{MSE}[MSE]{Mean Squared Error}

\newacro{MAC}[MAC]{Medium Access Control}

\newacro{NMSE}[NMSE]{Normalized Mean Squared Error}
\newacro{EM}[EM]{Electro-Magnetic}
\newacro{CSI}[CSI]{Channel State Information}
\newacro{MISO}[MISO]{Multiple-Input Single-Output}
\newacro{SISO}[MISO]{Single-Input Single-Output}
\newacro{NOMA}[NOMA]{Non-Orthogonal Multiple Access}
\newacro{OFDM}[OFDM]{Orthogonal Frequency-Division Multiplexing}
\newacro{QoS}[QoS]{Quality of Service}

\newacro{UAV}[UAV]{Unmanned Autonomous Vehicle}
\acrodefplural{UAV}[UAVs]{Unmanned Autonomous Vehicles}

\newacro{IoT}[IoT]{Internet of Things}
\newacro{LOS}[LOS]{Line Of Sight}
\newacro{NLOS}[NLOS]{Non-Line Of Sight}

\newacro{AoA}[AoA]{Angle of Arrival}
\acrodefplural{AoA}[AoAs]{Angles of Arrival}

\newacro{AoD}[AoD]{Angle of Departure}
\acrodefplural{AoD}[AoDs]{Angles of Departure}

\newacro{AWGN}[AWGN]{Additive White Gaussian Noise}
\newacro{AI}[AI]{Artificial Intelligence}
\newacro{SL}[SL]{Supervised Learning}
\newacro{RL}[RL]{Reinforcement Learning}
\newacro{DRL}[DRL]{Deep Reinforcement Learning}

\newacro{MDP}[MDP]{Markov Decision Process}
\acrodefplural{MDP}[MDPs]{Markov Decision Processes} 

\newacro{POMDP}[POMDP]{Partially Observable Markov Decision Process}
\acrodefplural{POMDP}[POMDPs]{Partially Observable Markov Decision Processes}

\newacro{DQN}[DQN]{Deep Q-Network}
\newacro{D$^3$QN}[D$^3$QN]{Dueling Double Deep Q Network}
\newacro{PG}[PG]{Policy Gradient}
\newacro{PPO}[PPO]{Proximal Policy Optimization}
\newacro{DDPG}[DDPG]{Deep Deterministic Policy Gradient}
\newacro{UCB}[UCB]{Upper Confidence Bound}
\newacro{GPU}[GPU]{Graphical Processing Unit}
\newacro{PDS}[PDS]{Post Decision State}
\newacro{TD}[TD]{Temporal Difference}
\newacro{MAML}[MAML]{Multi-Agent Reinforcement Learning}
\newacro{LSTM}[LSTM]{Long Short-Term Memory}

\newacro{EE}[EE]{Energy Efficiency}
\newacro{SE}[SE]{Spectral Efficiency}
\newacro{MIMO}[MIMO]{Multiple-Input Multiple-Output}

\newacro{IID}[IID]{Independent and Identically Distributed}

\newacro{MAB}[MAB]{Multi-Armed Bandits}
\newacro{CB}[CB]{Contextual Bandits}

\newacro{DRP}[DRP]{Deep Reward Prediction}

\newacro{DFT}[DFT]{Discrete Fourier Transform}
\newacro{DISAC}[DISAC]{Distributed Intelligent Integrated Sensing and Communication}
\newacro{ISAC}[ISAC]{Integrated Sensing and Communication}
\newacro{ML}[ML]{Machine Learning}

\newacro{SGD}[SGD]{Stochastic Gradient Descent}

\newacro{AI}[AI]{Artificial Intelligence}
\newacro{PoC}[PoC]{Proof of Concept}
\newacro{AGV}[AGV]{Automated Guided Vehicle}
\newacro{VRU}[VRU]{Vunlerable Road User}
\newacro{LIDAR}[LIDAR]{Light Detection and Ranging}
\newacro{V2X}[V2X]{Vehicle-to-Everything Communications}
\newacro{SLAM}[SLAM]{Simultaneous Localization and Mapping}

\newacro{V2V}[V2V]{Vehicle-to-Vehicle}
\newacro{RAN}[RAN]{Radio Access Network}
\newacro{O-RAN}[O-RAN]{Open Radio Access Network}

\newacro{SePF}[SePF]{Sensing Processing Function}
\newacro{SeMF}[SeMF]{Sensing Management Function}

\newacro{MIMO}[MIMO]{Multiple Input Multiple Output}
\newacro{D-MIMO}[D-MIMO]{Distributed MIMO}
\newacro{ELAA}[ELAA]{Extremely Large Aperture Antenna}
\newacro{KPI}[KPI]{Key Performance Indicator}
\newacro{KVI}[KVI]{Key Value Indicator}
\newacro{RAD}[RAD]{Range-Angle-Doppler}
\usepackage{upgreek}
\usepackage{algorithm}
\usepackage{algpseudocode}
\usepackage{placeins}
\usepackage[super]{nth}
\usepackage{dsfont}

\title{Distributed Intelligent Sensing and Communications for 6G: Architecture and Use Cases}
\author{Kyriakos Stylianopoulos, Giyyarpuram Madhusudan, Guillaume Jornod,\\ Sami Mekki, Francesca Costanzo, Hui Chen, Placido Mursia, Maurizio Crozzoli, \\Emilio Calvanese Strinati, George C. Alexandropoulos, and Henk Wymeersch}

\begin{document}

\maketitle

\begin{abstract}
The Distributed Intelligent Sensing and Communication (DISAC) framework redefines Integrated Sensing and Communication (ISAC) for 6G by leveraging distributed architectures to enhance scalability, adaptability, and resource efficiency. This paper presents key architectural enablers, including advanced data representation, seamless target handover, support for heterogeneous devices, and semantic integration.
Two use cases illustrate the transformative potential of DISAC: smart factory shop floors and Vulnerable Road User (VRU) protection at smart intersections. These scenarios demonstrate significant improvements in precision, safety, and operational efficiency compared to traditional ISAC systems.
The preliminary DISAC architecture incorporates intelligent data processing, distributed coordination, and emerging technologies such as Reconfigurable Intelligent Surfaces (RIS) to meet 6G’s stringent requirements. By addressing critical challenges in sensing accuracy, latency, and real-time decision-making, DISAC positions itself as a cornerstone for next-generation wireless networks, advancing innovation in dynamic and complex environments.
\end{abstract}
\begin{IEEEkeywords}
ISAC, 6G, Distributed Processing, Architecture, Use Cases. 
\end{IEEEkeywords}

\let\thefootnote\relax\footnotetext{This work has been supported by the SNS JU project 6G-DISAC under the EU’s Horizon Europe research and innovation program under Grant Agreement No 101139130.
K. Stylianopoulos and G. C. Alexandropoulos are with the Department of Informatics and Telecommunications, National and Kapodistrian University of Athens, 16122 Athens, Greece (e-mails: \{kstylianop, alexandg\}@di.uoa.gr).
G. Madhusudan is with the Technology \& Global Innovation-Orange Labs, Orange, France (e-mail: giyyarpuram.madhusudan@orange.com).
G. Jornod is with Bosch Corporate Research, Hildesheim, Germany (e-mail: Guillaume.Jornod@de.bosch.com).
S. Mekki is with Nokia Networks France (e-mail: sami.mekki@nokia.com).
F. Costanzo and E. Calvanese-Strinati are with  CEA-Leti, Université Grenoble Alpes Grenoble, France (e-mails: \{francesca.costanzo, emilio.calvanese-strinati\}@cea.fr).
H. Chen and H. Wymeersch are with the Department of Electrical Engineering, Chalmers University of Technology, 412 96 Gothenburg, Sweden (e-mail: \{hui.chen, henkw\}@chalmers.se).
P. Mursia is with NEC Laboratories Europe GmbH, 69115 Heidelberg, Germany (e-mail: placido.mursia@neclab.eu).
Maurizio Crozzoli is with TIM, Chief Mobile, Core \& Platforms Technology Office, Technology Innovation, Torino, Italy (e-mail: maurizio.crozzoli@telecomitalia.it).}

\section{Introduction}

In the emerging landscape of 6G networks, comprising enabling technologies play pivotal roles.
Among them, \ac{ISAC} has the potential to redefine the operation and capabilities of wireless environments \cite{9606831}.
Combining the traditionally separate domains of communication and sensing, \ac{ISAC} allows network entities to collect, process, and interpret environmental stimuli utilizing the same spectral, temporal, and computational resources used for communication.
This fusion supports tasks such as high-precision localization, object tracking, and environment mapping, which are critical for applications ranging from industrial automation to urban mobility, while accounting for dynamic reconfiguration and interoperation between the underlying subsystems \cite{10536135}.
As a result, \ac{ISAC} is instrumental in reducing hardware requirements and redundancy, organizing spectrum resources, and improving the system's energy efficiency \cite{3GPP_isac}.

Such ambitious objectives are often difficult to accommodate in the context of existing centralized systems, which may struggle with the scalability and adaptability requirements demanded by future applications.
Distributed systems are therefore purposely designed to mitigate such shortcomings through interconnected network devices and nodes that operate collaboratively.
This flexible architecture leverages edge computing capabilities for real-time responsiveness, and it, thus, not only enhances system reliability, robustness, and coverage, but also has the potential to significantly improve spatial resolution, sensing accuracy, data compression, and similar network objectives.
Under this viewpoint, distributed networks are particularly vital for dynamic scenarios of continuous area monitoring, such as \ac{V2V} coordination, environmental sensing, or smart city organization.

To fully realize their potential, \ac{DISAC} networks are mandated to overcome significant challenges~\cite{6G-DISAC-EuCNC24}.
A principal shortcoming lies in the lack of a comprehensive network architecture tailored to enable, capitalize upon, and enhance the heterogeneous nature of interconnected devices and emerging technologies.
In fact, current \ac{ISAC} and distributed frameworks do not always succeed in supporting diverse sensor types, dynamic operational conditions, and the integration of advanced technologies such as \ac{RIS}~\cite{BAL2024} and semantic-native processing.
In the absence of a unified design, the \ac{DISAC} transition faces the risk of operational fragmentation and application decoupling, leading to inefficient scalability and effectiveness.
Moreover, up to this point, the use cases utilized to illustrate the benefits of \ac{DISAC} are typically borrowed from existing proceedings \cite{3GPP_use_cases}, and therefore, they do not demonstrate the intricacy, potential, and quantifiable objectives of this rising paradigm.
The lack of dedicated \ac{DISAC} use cases hinders the validation of the aforementioned advantages in terms of system complexity, cost, and performance trade-offs, and presents a roadblock in providing a compelling narrative through practical applications for stakeholders and driving adoption across industries~\cite{6G-DISAC-magazine}.

Those gaps in the technology development motivate the current work in making the first steps toward presenting a preliminary \ac{DISAC} architecture and purposely designed use cases.
The architectural components are designed to highlight synergies among heterogeneous devices and the functionalities of communication and sensing, while each of the use cases exemplifies the technical capabilities and strength of this joint \ac{DISAC} treatment.
The proposed advancements present an intermediate step in the forthcoming 6G technology by providing comprehensive development and evaluation frameworks, but also laying the foundation for a new era of innovation, enabling 6G to truly transform how we connect, interact, and evolve in an increasingly complex digital world.

\section{Architectural Challenges and Enablers}\label{sec:enablers}

The \ac{DISAC} paradigm shift presents several critical challenges that must be addressed to fully realize its potential. First, finding a parsimonious data representation \textit{(C1)} is crucial to balance efficient processing and communication demands while maintaining high sensing performance. Second, ensuring the continuity of sensing services over large areas and extended timeframes \textit{(C2)} is essential, particularly for tracking mobile and passive targets seamlessly. Third, resource allocation must be managed intelligently to fulfill the diverse and dynamic requirements of \ac{DISAC} systems \textit{(C3)}. Additionally, meeting stringent \ac{KPI} and \ac{KVI} objectives \textit{(C4)} is imperative for validating system performance and utility. Finally, the integration of diverse network devices and elements with varying capabilities \textit{(C5)} poses architectural and operational challenges. These challenges are addressed through the architectural enablers summarized in Table \ref{table:challenges_enablers} and detailed in the following sections.


\begin{table}
\centering
\caption{DISAC Challenges and Enablers.}
\label{table:challenges_enablers}
\begin{tabular}{|p{0.7\linewidth}|p{0.2\linewidth}|}
\hline
\textbf{Enabler} & \textbf{Addresses Challenges} \\ \hline
Data Representation and Local vs Central Processing & C1, C4, C5 \\ \hline
Target Handover and Coordination & C2, C4 \\ \hline
Support for Heterogeneous Devices & C3, C4, C5 \\ \hline
Semantic Reasoning and Control & C1, C3, C4, C5 \\ \hline
\end{tabular}
\end{table}

\subsection{Data Representation and Local vs Central Processing}

\begin{figure}
    \centering
    \includegraphics[width=\linewidth]{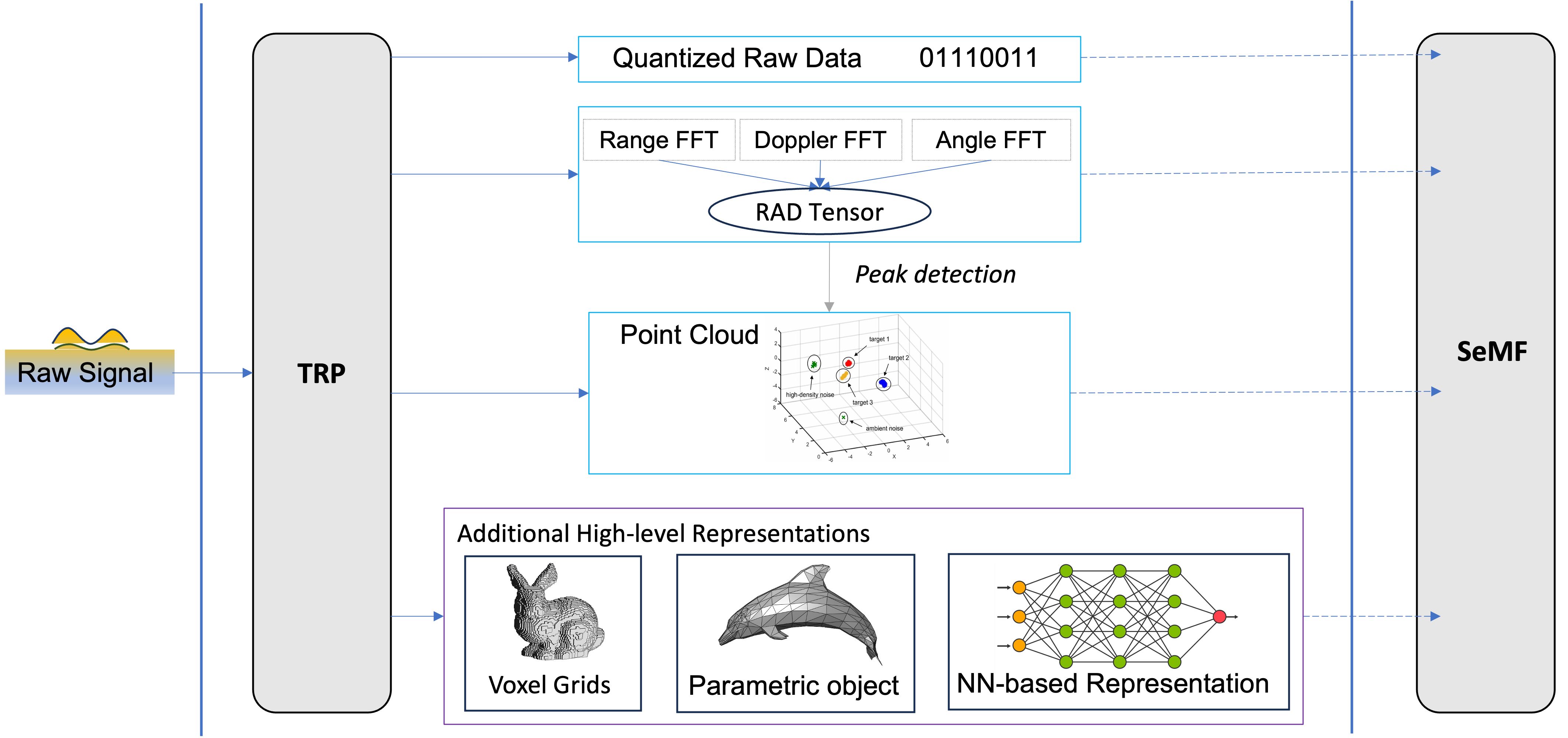}
    \caption{The process of information extraction from raw signals received at the Transmission Reception Point (TRP) to the Sensing Management Function (SeMF), highlighting different alternatives for data representations.}
    \label{fig:data-representation}
\end{figure}

In the \ac{DISAC} framework, efficient data representation and processing strategies are fundamental to achieving scalable and high-performing operations. The transition from raw sensing data to actionable insights involves multiple layers of abstraction, each tailored to optimize the trade-off between data transmission overhead and sensing accuracy (see Fig.~\ref{fig:data-representation}). Raw data, such as I/Q samples, can be voluminous, making its direct transmission impractical in distributed systems. Instead, higher-level representations, including  \ac{RAD} tensors, point clouds, and parametric object models, are leveraged to reduce communication requirements while preserving essential information~\cite{zhou2022towards}. Emerging techniques, such as deep learning-based representations and embeddings, further enhance data compression and semantic relevance~\cite{wang2022transformer}.

\begin{figure}
    \centering
    \includegraphics[width=\linewidth]{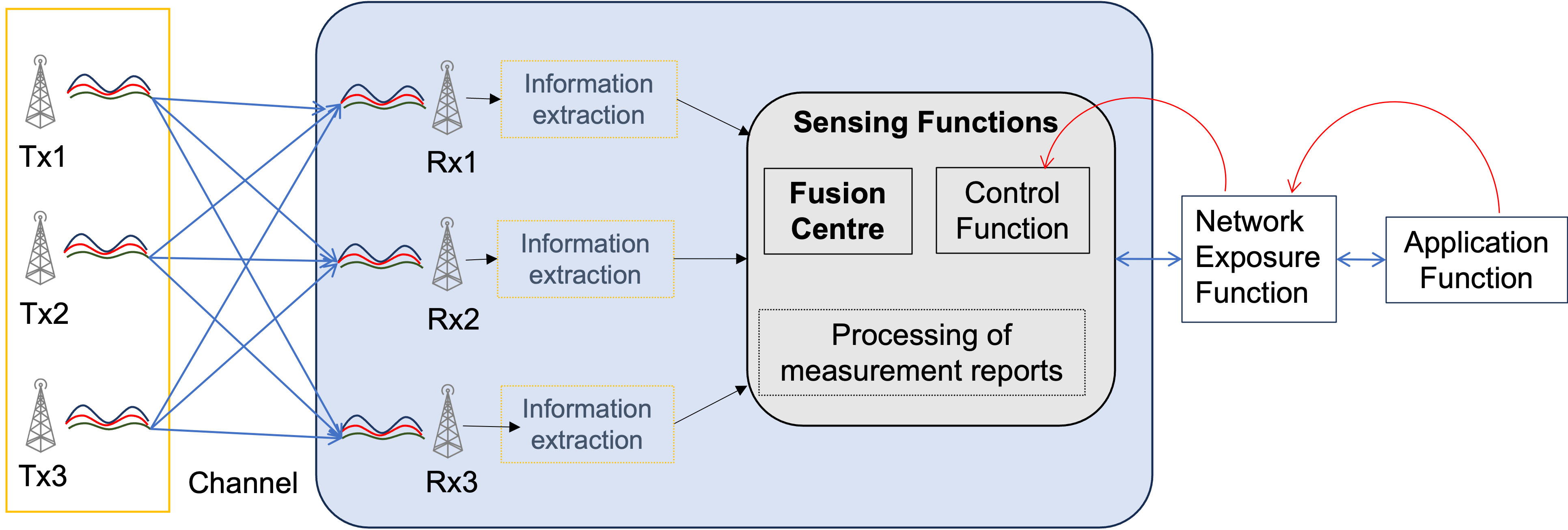}
    \caption{Distributed sensing signal processing and centralized fusion as a function interfacing with the core network.}
    \label{fig:signal-fusion}
\end{figure}

Processing in \ac{DISAC} systems must balance between fully centralized and fully local approaches. Centralized processing, where raw or minimally processed data is transmitted to a fusion center, can yield high accuracy through global optimization but is often constrained by latency and bandwidth limitations. Conversely, local processing at sensing nodes reduces communication load and enables real-time responses, though it may lead to suboptimal decisions due to limited context~\cite{zou2024distributed}. Hybrid architectures combining hierarchical or distributed processing are therefore preferred, where preliminary processing occurs locally, and higher-level decisions are made centrally or collaboratively, as depicted in Fig.~\ref{fig:signal-fusion}. This adaptability ensures the \ac{DISAC} framework meets diverse requirements across applications.

\subsection{Target Handover and Coordination}

\begin{figure}[t]
    \centering
    \includegraphics[width=0.7\linewidth]{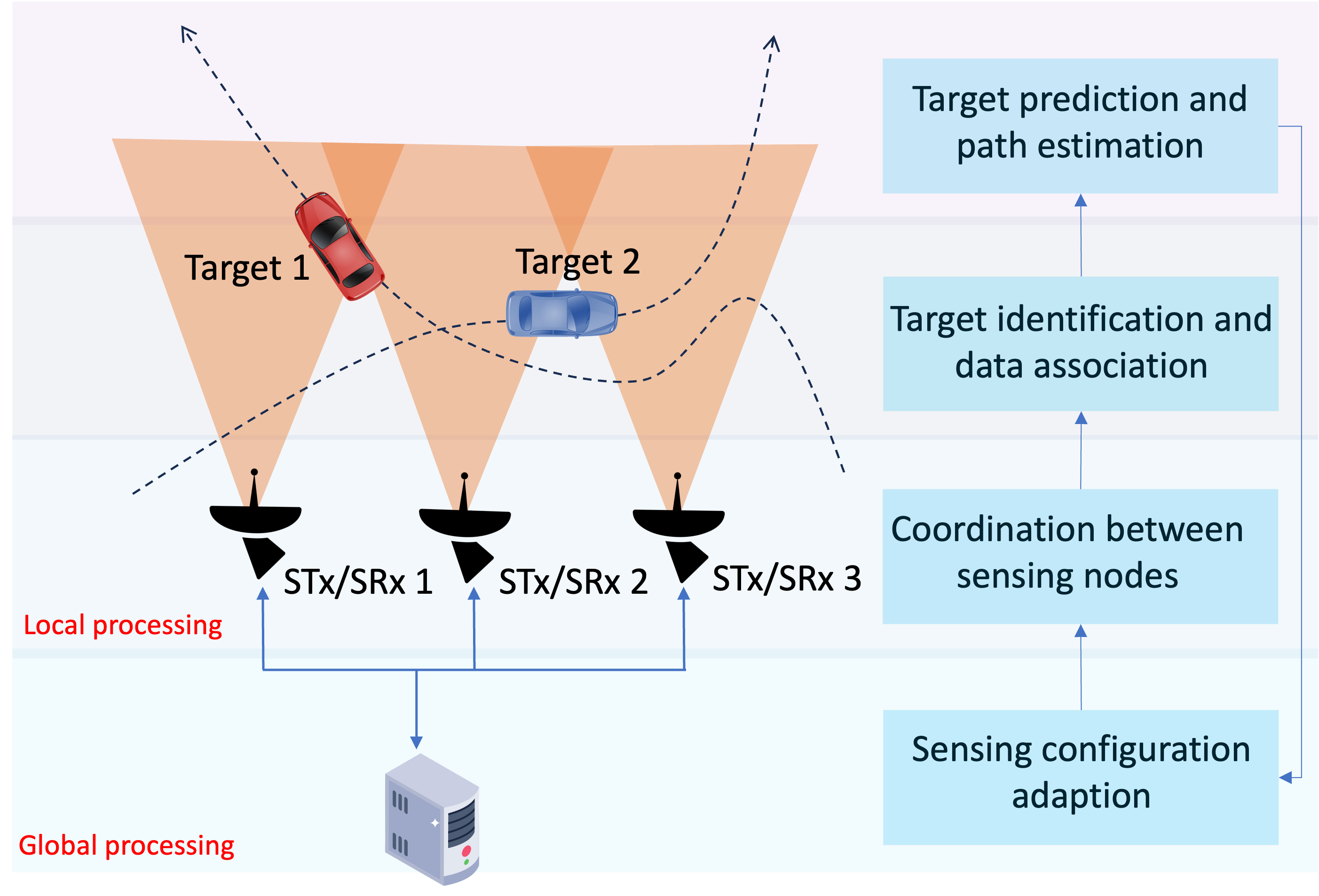}
    \caption{Illustration of multi-target tracking and handover, where multiple Sensing Transmitters (STx's) and Receivers (SRx's) cooperate to enlarge sensing areas through coordinate processing and re-configuration mechanisms.}
    \label{fig:handover}
\end{figure}

Efficient tracking and handover of targets across a distributed network of sensors is a cornerstone of the \ac{DISAC} framework. Traditional ISAC approaches often fail to maintain continuity when targets move beyond the field of view of individual sensors~\cite{ge2024target}. \ac{DISAC} addresses this through a coordinated network architecture that ensures seamless target handover, maintaining consistent identity and accurate tracking over large areas and extended durations. This capability is achieved by integrating multi-sensor data fusion, path prediction, and adaptive sensing configurations, as visually illustrated in Fig.~\ref{fig:handover}.

The architecture incorporates predictive models to anticipate target movement, allowing sensing nodes to prepare their resources in advance. This proactive approach minimizes sensing latency and resource contention, ensuring uninterrupted service. Additionally, a unified notion of target identifiers across sensing nodes is critical for effective data association and continuity. Coordination mechanisms between sensing nodes, facilitated by centralized or semi-centralized controllers, enable real-time adjustments to the sensing configurations, improving both accuracy and resource efficiency. Such robust coordination mechanisms are essential for applications like smart intersections and factory automation, where dynamic environments demand precision and adaptability.

\subsection{Support for Heterogeneous Devices}

The \ac{DISAC} framework embraces the heterogeneity of devices and their diverse capabilities, a necessity for addressing the varied requirements of 6G applications.
These devices range from low-power sensors to high-performance multi-antenna systems, each contributing unique strengths to the network the underlying architecture must seamlessly integrate, ensuring synchronized operations while addressing their inherent disparities in computational, energy, and communication capacities.
Massive \ac{MIMO}, \ac{D-MIMO}~\cite{D-MIMO}, holographic MIMO~\cite{HMIMO_survey}, and \ac{ELAA}s~\cite{ELAA}, alongside  \acp{RIS}~\cite{BAL2024} and full-duplex~\cite{FD_MIMO_ISAC_2024} exemplify advanced multi-antenna technologies that enhance sensing and communication precision, offering unparalleled flexibility in optimizing network performance~\cite{wang2024tutorial}. 
This diverse range of devices is characterized by different capabilities in terms of communication resources, RF configuration (e.g., beamforming), local processing and storage, and energy self-sufficiency.
Indeed, based on such characteristics, different levels of device autonomy and reconfigurability can be defined, ranging from non-reconfigurable devices that require no control to fully controllable and reconfigurable nodes and even autonomous entities with minimal control signaling.

\subsection{Semantic Reasoning and Control}

Semantic reasoning and control are transformative elements in the \ac{DISAC} framework, enabling a paradigm shift from traditional data-driven communication to meaning-driven operations. Semantics in \ac{DISAC} focuses on the extraction, representation, and utilization of meaningful information, reducing the need for raw data exchange while enhancing the system’s ability to make intelligent decisions. This approach is particularly valuable in distributed systems, where bandwidth constraints and latency requirements can hinder conventional data-intensive methods. Semantic communication integrates tightly with the foreseen \ac{DISAC} architecture, influencing waveform design, data representation, and network protocols. By prioritizing the transmission of semantically relevant information, the system reduces redundancy and ensures that resources are allocated to the most critical tasks. For example, semantic waveforms and the semantic \ac{RAN}  extend the \ac{DISAC} framework’s capabilities by embedding contextual information directly into the communication process. This integration not only improves resource efficiency but also enhances the system’s ability to adapt to dynamic environments. The semantic layer further enables advanced features such as intent-based network configurations and autonomous decision-making, solidifying \ac{DISAC} as a cornerstone of intelligent 6G networks.

\section{Use Cases}

The specific \ac{DISAC} use cases included in this work are selected primarily based on the following criteria: (i) \ac{DISAC} benefit, (ii) business potential, and (iii) demonstration potential. \ac{DISAC} benefit focuses on use cases requiring precise localization, detection, and identification of targets with high reliability, robustness, and resilience. Business potential identifies revenue sources for operators and other players in industry and smart city markets. Demonstration potential concerns the ability to illustrate and verify performance through demonstrators, leveraging expertise, existing \acp{PoC}, and testing facilities.

\subsection{Use Case 1: DISAC for Smart Factory Shop Floors}

The use case aims to enhance the operational capabilities of \acp{AGV} within a factory environment using specific frequency ranges of 6G technology.
\acp{AGV} benefit from additional information from the \ac{DISAC} system, particularly the extension of the field of view~\cite{AGV_challenges}.

\subsubsection{Use Case Implementation}

\acp{AGV} traverse the factory floor, communicating with the base station while sensing their surroundings. The \ac{DISAC} system optimizes both communication and sensing functions. The core components include a network controller, a base station, and the \acp{AGV}. The network controller manages communication and sensing tasks, ensuring optimal frequency resource allocation. The base station maintains robust links with all \acp{AGV}, facilitating real-time decision-making. \acp{AGV} perform real-time environmental mapping and navigation using mmWave bands for high-resolution sensing and Sub-6 GHz bands for stable communication: \acp{AGV} build a real-time map of the environment, crucial for operational efficiency. The implementation of radio \ac{SLAM} technology on \acp{AGV} exemplifies the \ac{DISAC} approach.

\subsubsection{User Story}

In a fully autonomous factory, all manufacturing processes are controlled and executed by machines and \acp{AGV}, without human intervention. The \acp{AGV} must maintain an accurate and up-to-date map of the factory floor, facilitated by high-resolution sensing capabilities. In a hybrid, dynamic scenario where humans and \acp{AGV} coexist, \acp{AGV} must navigate the factory floor while ensuring the safety of workers. This requires real-time sensing and adaptation to avoid collisions, supported by low-latency communication and robust safety protocols.

\subsubsection{Performance evaluation}
The performance of the system is evaluated based on the following indicators:
\begin{itemize}
    \item \textbf{Time Efficiency}: \ac{DISAC} reduces the time to complete environmental mapping with multiple \acp{AGV} working collaboratively.
    \item \textbf{Accuracy}: \ac{DISAC} reduces the error rate in the environmental map due to collaborative sensing and data integration.
    \item \textbf{Communication and Sensing Robustness and Resilience}: \ac{DISAC} reduces outage probability, data loss, and sensing service interruptions due to the distributed architecture and redundancy.
    \item \textbf{Real-time Data Processing and Decision Making}: \ac{DISAC} reduces latency from data collection to actionable decisions, indicating a more responsive system.
\end{itemize}

\subsection{Use Case 2: VRU Protection at a Smart Intersection}

\begin{figure}
    \centering
    \includegraphics[width=0.7\linewidth]{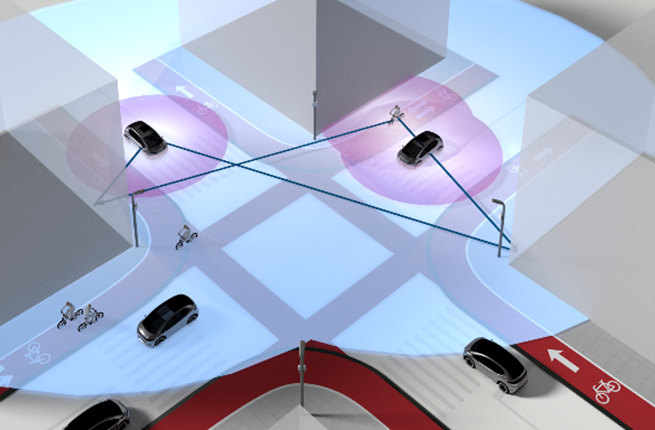}
    \caption{Use Case 2:  Traffic management for VRU protection at a smart intersection through by communication-based and multi-modal sensing.}
    \label{fig:UC2}
\end{figure}

Urban intersections are critical junctures where traffic flow intersects with the need to ensure the safety of \acp{VRU}, including pedestrians, cyclists, and others. This use case proposes \ac{DISAC} to enhance traffic management and improve \ac{VRU} safety at urban intersections (see Fig.~\ref{fig:UC2}).
By leveraging \ac{DISAC} with 6G networks~\cite{Veh_Ped_Comms}, \ac{IoT} devices, and \ac{AI} algorithms, the system aims to balance optimizing traffic flow and safeguarding \acp{VRU}. This is achieved through integrating communication-based sensing with other sensing modalities, ensuring real-time data collection and analysis~\cite{VRU_review}.

\subsubsection{User Story}
At an intersection where humans and vehicles (including autonomous vehicles) coexist, the system must ensure the safety of \acp{VRU} while maintaining efficient traffic flow. The system continuously senses the environment and adapts to traffic signals and vehicle movements in real-time to avoid collisions. Safety protocols prioritize \acp{VRU}' protection, and low-latency communication ensures prompt responses to dynamic changes of the environment, enhancing the situational awareness of drivers and \acp{VRU}.
\subsubsection{Use Case Implementation}

The system architecture integrates several key components, leveraging \ac{DISAC} alongside 6G network capabilities. \ac{IoT} sensors, including \ac{LIDAR}, radar, video cameras, and environmental sensors, are deployed at intersections to capture high-resolution data. The 6G network provides low-latency communication between sensors, vehicles, and the central traffic control system. The Traffic Control and Safety Centre acts as the hub for data aggregation, analysis, and decision-making, using \ac{AI} algorithms to detect safety risks and optimize traffic flow. It disseminates
safety alerts to connected vehicles and smartphones, enhancing situational awareness and coordinating rapid responses to incidents. DISAC enables distributed processing thus providing real-time performance for time critical updates..

\subsubsection{Performance evaluation}
The performance of the system is evaluated based on the following indicators:

\begin{itemize}
    \item \textbf{Time Efficiency}: \ac{DISAC} reduces the response time to traffic and \ac{VRU} movements with multiple sensors and devices working collaboratively.
    \item \textbf{Accuracy}: \ac{DISAC} reduces the error rate in the detection and localization of traffic and \acp{VRU} due to collaborative sensing and data integration.
    \item \textbf{Communication and Sensing Robustness and Resilience}: \ac{DISAC} reduces outage probability and data loss due to the distributed architecture and redundancy.
\end{itemize}

\subsection{DISAC Key Advantages over ISAC in Both Use Cases}

\ac{DISAC} offers several advantages over traditional \ac{ISAC} approaches. Firstly, \ac{DISAC} enables distributed sensing and mapping and fusion of information from different sensing modalities. While \ac{ISAC} might involve singular base stations or centralized systems performing sensing and communication tasks independently, \ac{DISAC} allows multiple base stations and devices to collaboratively perform these tasks. This distributed approach enhances signal coverage and accuracy, allowing for reliable detection, positioning, and classification of multiple targets in a scene, significantly speeding up the process and enhancing the detail and accuracy of the generated maps.
Secondly, \ac{DISAC} aims at improving efficiency and scalability. The efficiency of \ac{ISAC} in large or complex environments can be constrained by the limitations of individual base stations or centralized systems to cover the entire area quickly or accurately. \ac{DISAC} leverages the collective capabilities of multiple base stations and connected devices, allowing for more efficient coverage of large areas. This scalability is crucial in dynamic settings, such as urban or factory environments, where the environment can change rapidly, and new obstacles or areas of interest may emerge. 
Lastly, \ac{DISAC} enables robustness and redundancy. The reliance on centralized systems in \ac{ISAC} can introduce vulnerabilities, where the failure of a single component can significantly impact the system's overall performance. The distributed approach of \ac{DISAC} enhances system robustness and resilience. If one sensor, \ac{AGV}, or device encounters an issue, other ones can compensate, ensuring continuous operation and data collection. This redundancy is critical for maintaining high operational standards in both urban and industrial settings. The integration of 6G network infrastructure further enhances system robustness by ensuring reliable communication even in complex environments.

\section{Preliminary DISAC Architecture}

\begin{figure}
    \centering
    \includegraphics[width=0.9\linewidth]{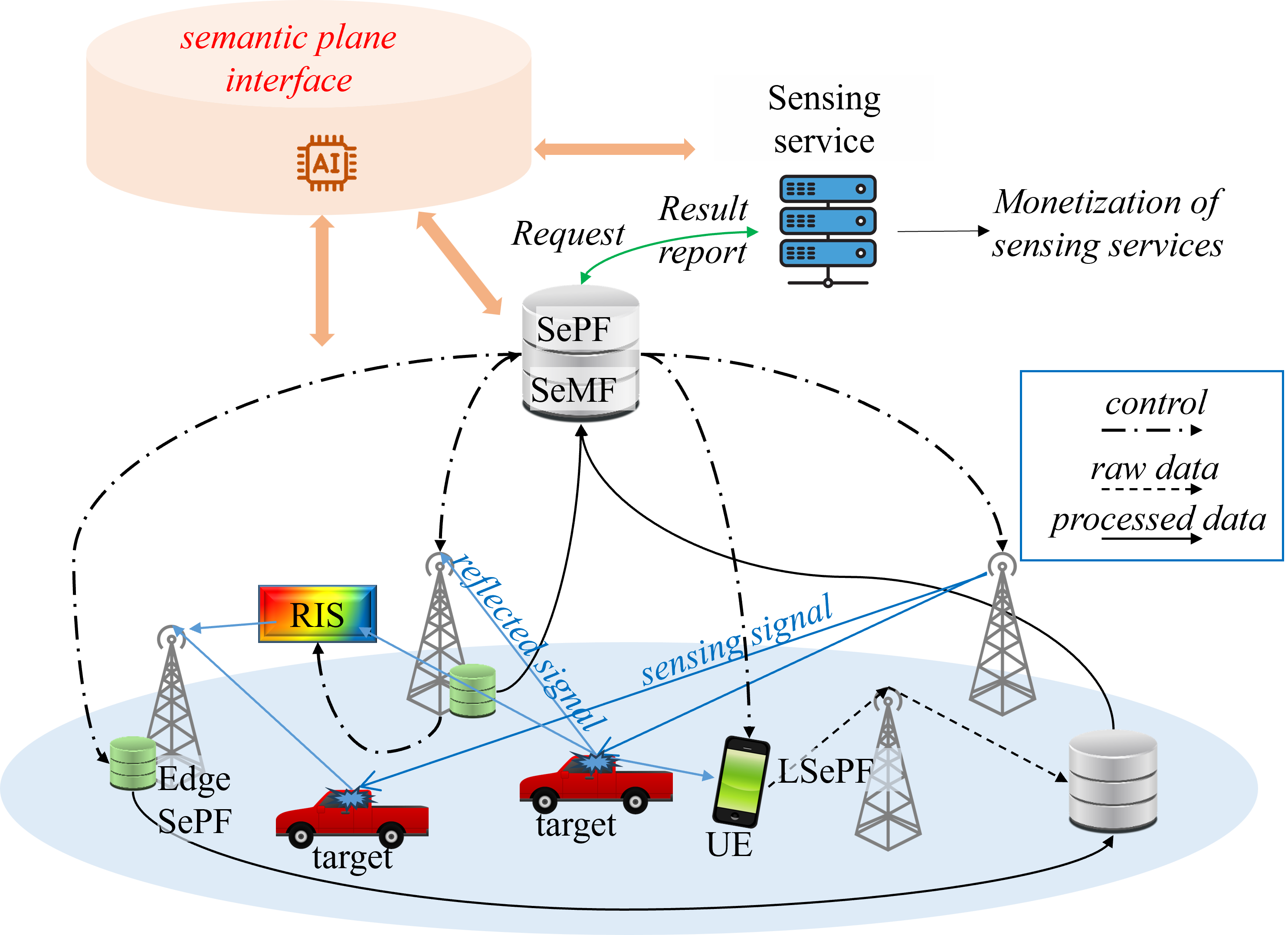}
    \caption{Overview of the proposed system architecture for distributed ISAC. Communication and sensing signals are collected and processed by the \acf{SePF}, upon being sent and reflected among the network infrastructure and the various connected User Equipment (UE) and passive targets. The \acf{SeMF} is responsible for interfacing with the sensing service.}
    \label{fig:arch}
\end{figure}

The preliminary \ac{DISAC} architecture represents a transformative approach for advancing 6G networks by combining the benefits of distributed sensing and communication within a unified framework. As illustrated in Fig.~\ref{fig:arch}, this architecture addresses the inherent challenges of transitioning from centralized ISAC to a distributed paradigm, focusing on scalability, adaptability, and efficiency. The proposed architecture integrates the advanced enablers of Section~\ref{sec:enablers} to meet the stringent requirements of diverse applications.
The \ac{DISAC} architecture is built on the foundation of distributed processing and information sharing, enabling multiple sensing and communication nodes to collaboratively achieve network goals. It is characterized by the ability to balance local and centralized data processing, allowing it to adapt to the demands of specific applications or use cases. By incorporating flexibility at its core, the architecture ensures robust performance in scenarios with diverse sensing and communication requirements, such as smart cities, autonomous mobility, and industrial automation.

\subsection{Data Processing and Representation}
A key feature of the preliminary architecture is its advanced data processing framework, which supports multiple levels of abstraction. Sensing data can be processed locally, hierarchically, or centrally, depending on the complexity of the task and the available network resources. Local processing allows nodes to perform initial tasks, such as clutter removal or preliminary target detection, reducing the volume of data transmitted to the central system. This minimizes communication overhead and latency. Conversely, central processing aggregates data from local nodes for global optimization and analysis.
The architecture also supports various data representation models, ranging from raw I/Q data to high-level semantic abstractions. These representations ensure efficient communication by reducing data redundancy while preserving essential information. Techniques such as parametric object models, point clouds, and deep learning-based representations enable compact yet meaningful data formats, facilitating their seamless integration into the distributed system.

\subsection{Target Tracking and Handover}

The architecture incorporates a robust framework for object tracking and target handover, essential for maintaining continuity in sensing services across large areas and extended periods. Target tracking relies on multi-sensor data fusion and predictive modeling to anticipate movements and allocate sensing resources effectively. The system assigns unique identifiers to targets, ensuring consistent tracking and reducing the likelihood of errors during handover between sensing nodes.
Handover is managed through coordinated mechanisms that involve real-time communication between nodes. As a target approaches the boundary of a node’s field of view, the architecture predicts its trajectory and prepares adjacent nodes to take over tracking responsibilities. This seamless process minimizes sensing gaps and ensures uninterrupted service, making the architecture particularly valuable for dynamic environments such as urban intersections or factory floors.

\subsection{Integration of Heterogeneous Devices}

The \ac{DISAC} architecture is designed to accommodate a wide range of devices with varying capabilities, from low-power sensors to high-performance base stations, ensuring efficient data handling and processing strategies while minimizing energy consumption across the network.
This heterogeneity enhances the flexibility and scalability of the system, allowing it to adapt to diverse applications.
Advanced multi-antenna technologies are integrated to improve spatial resolution and sensing accuracy.
Additionally, \acp{RIS} are employed to dynamically optimize signal propagation and sensing coverage with minimal installation and power consumption costs.
To manage the disparities in device capabilities, the architecture includes adaptive control mechanisms integrated into the \ac{SeMF}.
These mechanisms dynamically configure devices based on their roles, resource availability, and operational constraints, allowing support for fully controlled devices as well as (semi-) autonomous entities integrating local \acp{SePF}.
This ensures that all devices contribute effectively to the network, regardless of their individual limitations/capabilities, and allows the system to operate efficiently in resource-constrained environments.

\color{black}
\vspace{-0.1cm}
\subsection{Semantic Integration}

A distinctive feature of the \ac{DISAC} architecture is its incorporation of semantic processing, which enhances the system’s ability to understand and respond to contextual information. Semantic reasoning enables the network to extract meaningful insights from sensing data, reducing the need for exhaustive data exchange. This approach not only improves resource efficiency but also supports advanced features such as intent-based configurations and autonomous decision-making.
Semantic integration is achieved through specialized components such as the Semantic Sensing Configuration Assistant, which dynamically adjusts network parameters based on application requirements and environmental conditions. By embedding semantics into data representation and communication protocols, only relevant and actionable information is prioritized, streamlining operations and reducing computational overhead as well as power requirements in low-end devices~\cite{Stylianopoulos_Over_the_Air_RIS_DL}.

\vspace{-0.1cm}
\subsection{Deployment in O-RAN}

The \ac{O-RAN}~\cite{O-RAN} serves as a deployment framework for the \ac{DISAC} architecture, highlighting its compatibility with emerging industry standards. 
\ac{O-RAN}’s modular and open design allows for seamless integration of \ac{DISAC} components, facilitating the deployment of distributed processing and semantic-based communication strategies. The data variation and richness provided by \ac{O-RAN}, originating from diverse network sources, 
enhance distributed sensing by enabling more comprehensive environmental awareness and adaptive decision-making. 
This diversity improves the accuracy and robustness of semantic reasoning, target handover, and cooperative sensing, making the network more resilient and intelligent.
At the same time, potential enhancements are envisioned toward the integration of the different types of devices such as \acp{RIS}, building upon emerging architectural proposals~\cite{rise-6g-eurasip}.

\vspace{-0.1cm}
\section{Conclusion and Outlook}
This paper has presented a comprehensive overview of the \ac{DISAC} framework from the architectural perspective, with respect to key emerging supporting technologies and dedicated use cases.
Two main use cases have been showcased, namely \ac{AGV} coordination in smart factories and \ac{VRU} protection at smart intersections, to the purpose of providing benchmark scenarios to motivate, develop, evaluate, and illustrate the need for \ac{ISAC} and the benefit of the \ac{DISAC} approach.
The preliminary version of the proposed architecture highlights the importance of local and centralized \ac{SePF} entities that process the information bearing signals collected via diverse devices and different \ac{ISAC} procedures before forwarding them to the orchestrating \ac{SeMF} component interfacing with the core-network.
The comprehensive view of the architecture is designed to accommodate the architectural enablers presented earlier, while being compatible with modular \ac{O-RAN} architectures.
Overall, \ac{DISAC} offers significant advancements in scalability, adaptability, and resource efficiency, establishing it as a cornerstone for next-generation wireless systems.

As further steps, dedicated \ac{ISAC} algorithms and methodologies will have to be implemented that furthermore take account of the functional components and inherent trade-offs of local vs centralized sensing data processing and transmission.
To this end, orchestration approaches will need to be developed that leverage novel protocols for efficient allocation of radio and computational resources.
The development of a refined version of the architecture, where the underlying functional components, employed devices, and operations fully integrate with \ac{O-RAN} is of particular importance, while the implementation of the proposed use cases via the means of realistic simulations and \acp{PoC} can pave the way for comprehensive validation and dissemination frameworks.

\FloatBarrier
\bibliographystyle{IEEEtran}
\vspace{-0.1cm}
\bibliography{references}

\end{document}